\documentclass[useAMS,usenatbib]{mn2e}

\usepackage{graphicx}
\usepackage{amssymb}
\usepackage{amsmath}

\newcommand{\apj}{ApJ}
\newcommand{\apjs}{ApJS}
\newcommand{\apjl}{ApJ}
\newcommand{\mnras}{MNRAS}
\newcommand{\aap}{A\&A}

\newcommand{\solphys}{Sol. Phys.}

\def\<<{{\ll}}
\def\>>{{\gg}}

\def\spose#1{\hbox to 0pt{#1\hss}}
\def\ltwig{\mathrel{\spose{\lower 3pt\hbox{$\mathchar"218$}}
     \raise 2.0pt\hbox{$\mathchar"13C$}}}
\def\gtwig{\mathrel{\spose{\lower 3pt\hbox{$\mathchar"218$}}
     \raise 2.0pt\hbox{$\mathchar"13E$}}}
\def\+/-{{\pm}}
\def\=={{\equiv}}
\def\Beq{B_{eq}}
\def\Rstar{R_{\ast}}

\def\Mdot{\dot M}

\def\yr{{\rm yr}}

\def\solar{\odot}
\def\Msun{M_{\solar}}
\def\msbyr{\Msun/\yr}

\def\vinf{v_\infty}

\newcommand{\beq}{\begin{equation}}
\newcommand{\eeq}{\end{equation}}
\newcommand{\beqa}{\begin{eqnarray}}
\newcommand{\eeqa}{\end{eqnarray}}

\begin{document}

\title[Angular Momentum Loss and Rotational Spindown]
   {Dynamical Simulations of Magnetically Channeled Line-Driven Stellar Winds:
\\  III. Angular Momentum Loss and Rotational Spindown}
 \author[ud-Doula et al.]
 {Asif ud-Doula$^1$\thanks{Email: uddoula@morrisville.edu},
 Stanley P. Owocki $^2$ and Richard H.D. Townsend$^3$ \\
 $^1$ Department of Physics, Morrisville State College,
 Morrisville, NY 13408, USA.\\
 $^2$ Bartol Research Institute,
     University of Delaware,
     Newark, DE 19716, USA.\\
 $^3$ Department of Astronomy, University of Wisconsin-Madison,
 5534 Sterling Hall, 475 N Charter Street, Madison, WI 53706, USA.
}
\date{\today}

\maketitle

\begin{abstract}
We examine the angular momentum loss and associated rotational
spindown for magnetic hot stars with a line-driven stellar wind
and a rotation-aligned dipole magnetic field.
Our analysis here is based on our previous 2-D numerical MHD simulation
study that examines the interplay among wind, field, and rotation
as a function of two dimensionless parameters,
one characterizing the wind magnetic confinement
($\eta_{\ast} \equiv B_{eq}^{2} R_{\ast}^{2}/{\dot M} v_{\infty}$),
and the other the ratio ($W \equiv V_{rot}/V_{orb}$) of stellar rotation to
critical (orbital) speed.
We compare and contrast the 2-D, time variable angular momentum loss of
this dipole model of a hot-star wind with the classical 1-D steady-state
analysis by Weber and Davis (WD), who used an idealized monopole field
to model the angular momentum loss in the solar wind.
Despite the differences, we find that the total angular momentum
loss ${\dot J}$ averaged over both solid angle and time follows closely the
general WD scaling ${\dot J} = (2/3) {\dot M} \Omega R_{A}^{2}$,
where ${\dot M}$ is the mass loss rate, $\Omega$ is the stellar angular
velocity, and $R_{A}$ is a characteristic Alfv\'{e}n radius.
However, a key distinction here is that for a dipole field,
this Alfv\'{e}n radius has a strong-field scaling
$R_{A}/R_{\ast} \approx \eta_{\ast}^{1/4}$,
instead of the scaling $R_{A}/R_{\ast} \sim \sqrt{\eta_{\ast}}$ for a monopole
field.
This leads to a slower stellar spindown time that in the dipole case scales
as $\tau_{spin} = \tau_{mass}\, 1.5k/\sqrt{\eta_{\ast}}$,
where $\tau_{mass} \equiv M/{\dot M}$ is the characteristic mass loss
time, and $k$ is the dimensionless factor for stellar moment of inertia.
The full numerical scaling relation we cite gives typical spindown
times of order 1~Myr for several known magnetic massive stars.

\end{abstract}

\section{Introduction}

In recent years, improvements in spectropolarimetry  have made it
possible to detect moderate to strong ($10^{2}-10^{4}$G) magnetic
fields in a growing number of hot, luminous, massive stars of spectral
type O and B \citep[e.g.,][]{Don2002}.
The high luminosity of such stars drives a strong stellar wind
through line scattering of the star's continuum radiation
(\citealt{Cas1975}, hereafter CAK).
The first two papers in this series  focus on developing
numerical magnetohydrodynamics (MHD) simulations of the confinement and
channeling of this stellar wind outflow by a dipole magnetic field at
the stellar surface.
For models without rotation, Paper 1 \citep{udDOwo2002}
showed that  the overall effect of the field depends on a
single ``wind magnetic confinement parameter'' $\eta_{\ast}$
(defined in eqn.\ (\ref{esdef}) below)
that characterizes the ratio of magnetic to wind energy density near
the stellar surface.
Paper 2 \citep{udDOwo2008}
extended this study to include field-aligned
rotation, examining a wide range of magnetic confinement parameters
($\eta_{\ast}$ from near unity up to 1000) in stars with equatorial speeds
that are  a substantial fraction ($W=$1/4, 1/2) of the critical (orbital) speed.
Paper 2 focussed mainly on formation and disruption of equatorial
rigid-body disks.
The present study utilizes the same 2-parameter MHD simulation study
to examine the angular momentum loss and associated stellar spindown
for this case of wind outflow from a hot star with a rotation-aligned dipole.

Most of the previous literature on magnetic wind spindown has focussed
on cool, solar-type stars, for which the wind is driven by the high
gas pressure of a corona heated to more than a million K
[see, e.g., reviews by \citet{Mes1968a,Mes1968b,Mes1984}  and
\citet{MesSpr1987}; \citet{TouPri1992}].
The mechanical energy to heat the corona is thought to originate from the
strong convection in the hydrogen recombination layers below the stellar surface.
Moreover, the interaction of this convection with stellar rotation is
understood to drive a dynamo that generates a complex stellar magnetic field
and activity cycle.
Much of the emphasis in studying cool-star spindown has thus focussed on
the feedback of faster rotation in generating both a stronger
field and more mechanical heating to drive a stronger coronal wind,
which then act together to give a more abrupt spindown for younger, more-rapidly
rotating stars.
For a middle-age star like the sun, which is roughly halfway through
its expected 10~Byr main-sequence lifetime, the rotation speed is thus
quite slow, only about 2~km/s at the solar equator,
with an associated rotation period of about 27~days.

For massive, hot, luminous stars with radiatively driven winds, the direct
study and modeling of wind magnetic spindown is more limited.
This is partly due to a general expectation that the absence of a
hydrogen recombination convection zone means that hot stars should not
have the magnetic dynamo activity cycles of cooler stars.
Nonetheless, as noted above, spectropolarimetric observations have
directly detected large-scale fields in a growing number of O-type (currently 3)
and early-B-type (ca. 2 dozen) stars, often characterized by a more or
less constant dipole that is tilted relative to the stellar rotation axis.
This steady nature and large-scale contrast with the active and complex magnetic
activity cycle of cool stars, and suggests a primordial fossil origin
instead of active dynamo generation.
However there have been models based on active generation in the
convective core (\citealt{MacCas2003}; \citealt{MacCha1999})
or envelope \citep{MulMac2001,MulMac2005}.

All the directly detected, oblique dipoles with strong fields also
exhibit a clear periodic rotational modulation in circumstellar signatures
like X-ray emission and in wind signatures like the UV  P-Cygni line profiles.
But weaker, as-yet undetected, smaller-scale fields might also be one cause
of less regular wind variations,  such as the episodic
discrete absorption components (DACs) commonly seen in the absorption troughs
of P-Cygni profiles.

Compared to cooler stars, the rotation in early-type stars is quite fast
with inferred periods typically one to several days, and projected
surface rotation speeds ($V\sin i$) of order one or two {\em hundred}~km/s.
This has even been used to argue for a {\em lack} of wind magnetic spindown,
and thus for a lack of a dynamically significant magnetic field.
Based on estimates by \citet[][hereafter FM84]{FriMac1984}
for the dependence of  spindown time on hot-star mass loss rate
and magnetic field strength, \citet{Mac1992} argued that
magnetic fields in massive stars with still-rapid rotation must generally
be less than about 100~G.

In the context of the present paper, a key issue for this FM84
analysis of wind magnetic spindown in hot stars is that it is based on
the idealization
-- first introduced in the seminal paper by \citet[][hereafter WD67]{WebDav1967}
on spindown from the solar wind --
that the radial field at the stellar surface
can be described as a simple {\em monopole}.
Although magnetic fields can never be actual monopoles,
this idealization greatly simplifies the analysis, making tractable a
quasi-analytic, 1-D formulation.
Moreover, for the sun it is somewhat justified by the inference that,
beyond a few solar radii from the solar surface, the outward expansion
of the solar wind pulls the sun's complex multipole field
into an open radial configuration,
roughly characterized by a ``split monopole'', with opposite
polarity on the northern vs.\ southern sides of a heliospheric
current sheet.

However, as noted above, the fields detected in hot stars
are often inferred to be dominated by a large-scale, {\em dipole} component.
For such cases, it is not clear how applicable the monopole-based
analysis of WD67 and FM84 should be for estimating spindown rates.
Indeed, while some magnetic O-stars have slow rotation periods
(e.g.\ 15~days for $\theta^{1}$~Ori~C, and ca. 500~d for HD191612;
\citet{Don2006b}),
several early-B stars with very strong (many kG) magnetic fields
still retain a quite rapid rotation
(e.g.\ period of 1.2~days for $\sigma~$Ori~E, see \citealt{GroHun1982}).

The central purpose of this paper is to use our previous MHD simulation
parameter study to examine the wind magnetic spindown of massive stars
with a rotation-aligned {\em dipole} field.
Compared to the 1-D  semi-analytic studies for an idealized
monopole field, the numerical simulations of dipole fields here requires
a second spatial dimension (latitude), but the restriction to field-aligned
rotation allows us to retain a 2-D axisymmetry in which the variations in
azimuth are ignorable.
However, our `2.5-D' formulation still retains the crucial azimuthal components
of the flow velocity and magnetic field.

As discussed in detail in section 3 below, this
parameter study shows that the loss of angular momentum for this
dipole case is a highly complex, time-variable process, characterized
by a quasi-regular cycle of build-up and release of angular momentum
stored in the circumstellar field and gas.
But quite remarkably, the time- and angle-averaged total angular
momentum loss follows a simple scaling rule that is quite analogous to
that derived  by WD67, with however a key dipole modification in the
scaling of the associated Alfv\'{e}n radius relative to that applicable
to the simple WD67 monopole model.
As discussed in section 4, this leads to a substantial general reduction
in the spindown rate for this dipole case relative to that derived by
FM84 for hot-stars using a monopole model.
A concluding section 5 summarizes the results and outlines directions
for future work.
To lay the groundwork for interpreting the detailed numerical simulation
results, the next section gives some essential background on
the basic analytic scaling laws for  angular momentum loss in the WD67
monopole model and some minor variants.

\section{Background}

\subsection{The Weber \& Davis monopole model for the solar wind}

An outflowing wind carries away
angular momentum and thus spins down the stellar rotation.
Winds with magnetic fields exert a braking torque
that is significantly larger than for non-magnetic cases,
due to the larger lever arm of magnetic field lines
that extend outward from the stellar surface.

A seminal analysis of this process was carried out by
WD67, who modelled the angular momentum
loss of the solar wind for the idealized case of a simple {\em monopole}
magnetic field from the solar surface.
In terms of the surface angular velocity $\Omega$ and wind mass loss rate
${\dot M}$,
a key result is that the {\em total} angular momentum loss rate scales as
\beq
{\dot J}_{}
= {2 \over 3}\, \Mdot \Omega R_A^2 \, ,
\label{Jtot-wd}
\eeq
with $R_A$ the Alfv\'{e}n radius, defined by where the radial
components of the field and flow have equal energy density.
This can be intuitively interpreted as the angular momentum loss that
{\em would} occur {\em if} the gas were kept in rigid-body rotation up to
$R_{A}$, and then effectively released.
But while helpful as a kind of mnemonic, it is not literally the case,
since in fact, as WD67 emphasized (and is discussed further below),
most of the angular momentum is actually lost via Poynting stresses of
the magnetic field, and {\em not} by the gas itself.

For any radius $r$, the energy density ratio between radial field and flow
is given by
\beq
\eta(r) \equiv \frac{B_{r}^{2}/8\pi}{\rho v_{r}^{2}/2}
= \left ( \frac{V_{A}}{v_{r}} \right )^{2}
= M_{A}^{-2}
\, .
\label{eta-eqn}
\eeq
The latter equalities emphasize this energy ratio can also be cast as the
inverse square of the Alfv\'{e}nic Mach number,
$M_{A} \equiv v_{r}/V_{A}$, where the radial Alfv\'{e}n speed,
$V_{A} \equiv B_{r}/\sqrt{4 \pi \rho}$, with $\rho$ the wind mass
density.
The Alfv\'{e}n radius is then defined implicitly by
$\eta(R_{A})\equiv 1$.

Using detailed flow solutions of the equations for a gas-pressure-driven
solar wind, together with {\em in situ} measurements of the radial magnetic field
near earth's orbit, WD67 estimated the solar wind  Alfv\'{e}n radius
to be $R_{A} \approx 24.3 R_{\odot}$.
With this extended moment arm, even the quite low solar wind mass loss rate
implies a substantial spindown over the solar lifetime, providing a
possible explanation of the slow solar rotation.
Applications to other solar-type stars \citep[e.g.][]
{Mes1968a,Mes1968b,MesSpr1987,Kaw1988,TouPri1992}
have largely focussed on the potential feedback of rotation on the field strength
and mass loss rate.

But in the present context of hot-star winds, for which the
mass loss rate is set near the surface by the physics of radiative
driving, we can derive approximate {\em explicit} expressions in terms
of fixed  values for the equatorial field strength $\Beq$ at the
surface radius $R_{\ast}$, and for the wind mass loss rate ${\dot M}$ and
terminal flow speed $\vinf$.
Specifically, following papers 1 and 2, if we define here a wind
{\em magnetic confinement parameter},
\beq
\eta_{\ast} \equiv \frac {\Beq^2 \, \Rstar^2} {\dot{M} \, \vinf}
\, ,
\label{esdef}
\eeq
then we can write the energy density ratio in the form
\beq
\eta(r) =  \eta_{\ast}
\,
\left [ \frac{r}{\Rstar} \right ]^{2-2q}
\, \frac{\vinf}{v_{r}(r)}
\approx \frac{\eta_{\ast}}{(r/\Rstar)^{2}(1- \Rstar/r)^{\beta}}
\, ,
\label{eta-beta}
\eeq
where $q$ is the power-law exponent for radial decline of
the assumed magnetic field,
and the latter equality assumes now a monopole field ($q=2$),
together with a canonical `beta' velocity law with index $\beta$,
and terminal speed $\vinf$.

For the monopole case and velocity index of either $\beta =1$ or
$\beta=2$, an explicit expression for the Alfv\'{e}n radius can be
found from solution of a simple quadratic equation, yielding
\beq
\frac{ R_{A}}{ R_{\ast}} = 1/2 + \sqrt{\eta_{\ast} + 1/4} ~~ ; ~~
\beta = 1
\label{rawd-beta1}
\eeq
and
\beq
\frac{ R_{A}}{ R_{\ast}} = 1 + \sqrt{\eta_{\ast}} ~~ ; ~~
\beta = 2
\, .
\label{rawd-beta2}
\eeq
For weak confinement, $\eta_{\ast} \ll 1$, we find
$R_{\rm{A}} \rightarrow \Rstar$,
while for strong confinement, $\eta_{\ast} \gg 1$, we obtain
$R_{\rm{A}} \rightarrow \sqrt{\eta_{\ast}} \Rstar$.
Application in eqn.~(\ref{Jtot-wd}) then gives an explicit expression
for the angular momentum loss rate.

Note that in the strong magnetic
confinement limit $\eta_{\ast} \gg 1$, the scaling
$R_{A} \sim \sqrt{\eta_{\ast}}$ implies that the angular momentum loss
for this monopole model becomes {\em independent} of the mass loss rate,
\beq
{\dot J} \approx \frac{2}{3} \, \Mdot \Omega R_\ast^2  \eta_{\ast}
 = \frac{2}{3} \, \frac{\Omega R_\ast}{\vinf} \, \Beq^{2} R_{\ast}^{3}
~~ ; ~~
\eta_{\ast} \gg 1  \, .
\eeq
For a star of moment of inertia $I= k M_{} R_{\ast}^{2}$ (with
typically $k \approx 0.1$),
the associated characteristic {\em spindown time} for stellar rotation is
\beq
\tau_{spin} \equiv \frac{J}{{\dot J}}
\approx \frac{ k M_{} \Omega R_{\ast}^{2}}{\frac{2}{3}\Mdot \Omega R_A^2}
= \frac{\frac{3}{2} k}{\eta_{\ast}} \, \tau_{mass}
= \frac{\frac{3}{2}k M_{} \vinf}{\Beq^{2} R_{\ast}^{2}}
\, ,
\label{tauspin-wd}
\eeq
where the third equality gives a scaling in terms of a characteristic
mass loss time, $\tau_{mass} \equiv M_{}/{\dot M}$, and the last
equality  gives the mass-loss--independent scaling. Note that
although we derived this scaling in the context of
line-driven hot-star winds, we did not make any explicit
assumptions about the wind-driving mechanism.
As such, we can apply this even to gas-pressure driven winds.
In particular, for the solar wind, {\it in situ} measurements near 1~au give a speed
$\vinf \approx 400 ~$km/s and radial field strength $B_{au} \approx 5
\times 10^{-5}$~G, translating to a monopole field strength
$\Beq \approx B_{au} (au/R_{\odot})^{2} \approx 2.3~$G at the
solar surface.
Using $k = 0.059$, eqn. (\ref{tauspin-wd}) then gives a solar
spindown time of ca.\ 8.6~Byr, comparable to the
solar age of ca.\ 5~Byr.

\subsection{Angular momentum loss from gas vs. magnetic field}

Let us next consider  general expressions for angular momentum
loss, comparing the contribution due to the gas vs.\ the magnetic field.
For spherical coordinates representing
radius $r$, co-latitude $\theta$, and azimuth $\phi$,
let subscripts denote associated components of the
vector velocity $(v_{r}, v_{\theta}, v_{\phi})$ or
magnetic field, $(B_{r}, B_{\theta}, B_{\phi})$.
We are interested in the angular momentum about the rotation
axis.
For the gas, the associated angular momentum per unit mass is given by
the azimuthal speed times the distance to the rotation axis,
$v_{\phi} r \sin \theta$.
Multiplying this by the mass flux density $\rho v_{r}$ then gives the
angular momentum flux within an element of area $r^{2} d\phi d\mu$
(defining $\mu \equiv \cos \theta$).
Upon integration over azimuth (assuming axisymmetry), we obtain the latitudinal distribution
of gas angular momentum loss at any radius and colatitude,
\beq
\frac{d{\dot J}_{gas}}{d\mu} = {\dot m} v_{\phi} r
\frac{\sin \theta}{2}
\, ,
\label{dJgdmu}
\eeq
where ${\dot m} \equiv 4 \pi \rho v_{r} r^{2}$ gives the local mass loss
rate (which in general could vary in latitude, radius, or time).

For the magnetic field, the angular momentum loss is proportional to
the $r,\phi$ component of the Maxwell stress tensor,
\beq
T_{r\phi}=-\frac {B_r B_\phi}{4 \pi}
\, ,
\label{trphi}
\eeq
which represents the radial flux density of azimuthal momentum.
As before, multiplying by the axial distance $r \sin \theta$
converts this into an associated angular momentum flux, which upon
azimuthal integration gives the latitudinal distribution of magnetic
angular momentum loss,
\beq
\frac{d{\dot J}_{mag}}{d\mu} = - r^{2} B_{r} B_{\phi} r
\frac{\sin \theta}{2}
\, .
\label{dJmdmu}
\eeq

\subsection{${\dot J}_{gas}$ vs. ${\dot J}_{mag}$ in the
Weber-Davis model}

These expressions for loss of rotational angular momentum
apply for any general magnetic field, including the
rotation-aligned dipole model discussed in detail in \S 3.
But to illustrate some characteristic properties, let us first examine
them for the simple WD67 monopole field model.
In this case,  both $v_{\phi}$ and $B_{\phi}$ scale in proportion
to $\sin \theta$, giving an overall latitudinal dependence with
$\sin^{2} \theta = 1-\mu^{2}$.
For a slow rotation case like the sun, $B_{r}$, $v_{r}$, $\rho$ and
${\dot m}$ are  otherwise largely independent of
latitude \footnote{However, in monopoles winds with more rapid
rotation,
the  field and outflow both tend to become deflected toward the rotation pole;
see \citet{SueNer1975} and \citet{WasShi1993}.}.
As such, latitudinal integration (from $\mu =-1$ to $\mu=+1$)
gives an overall angular momentum loss that is just a factor 2/3
smaller than computed from the WD67 equatorial analysis,
\beq
{\dot J}_{}
= \frac{2}{3} {\dot J}_{eq}
= \frac{2}{3} \left [
{\dot m} v_{\phi} r
- r^{2} B_{r} B_{\phi} r
\right ]_{eq}
\, .
\label{Jdot-gas+mag}
\eeq

Since under the ``frozen-flux'' condition (applicable to
ideal MHD) the local velocity vector
is parallel to the local field,
we can relate the equatorial $B_{\phi}$ and $v_{\phi}$ through,
\beq
\frac{B_{\phi}} {B_{r}} = \frac{\Omega r - v_{\phi}}{v_{r}}
\, .
\label{Bphi-ff}
\eeq
Defining then an equatorial specific angular momentum
$j_{eq} \equiv {\dot J}_{eq}/{\dot m}$, we find that combining eqns.
(\ref{Jdot-gas+mag}) and (\ref{Bphi-ff})
gives for the gas specific angular momentum,
\beq
j_{gas} \equiv r v_{\phi} =
\frac
{j_{eq} \, M_{A}^{2} - \Omega r^{2}}
{M_{A}^{2} - 1}
\, .
\label{jg-wd}
\eeq
At the Alfv\'{e}n radius $R_{A}$, where the Alfv\'{e}nic Mach number
$M_{A} = 1$, the denominator vanishes;
ensuring continuity thus requires that the numerator also must vanish
at this point, which implies $j_{eq} = \Omega {R_{A}}^{2}$.
This thus provides the basis for the key WD67 scaling cited in eqn. (\ref{Jtot-wd}).

The fraction of angular momentum carried by the gas at any radius is
then given by
\beq
\frac{j_{gas}}{j_{eq}} = \frac{r v_{\phi}}{\Omega {R_{A}}^{2}}
= \frac
{1-v_{rA}/v_{r}}
{1-v_{rA}{R_{A}}^{2}/v_{r} r^{2}}
\, ,
\label{jgfrac-wd}
\eeq
where $v_{rA} \equiv v_{r} (R_{A})$.
In the spherical expansion of the WD67 model, a similar 2/3
latitudinal correction
applies to both the gas and total angular momentum, and so eqn.\
(\ref{jgfrac-wd}) also gives the spherically averaged gas angular
momentum fraction, ${\dot J}_{gas}/{\dot J} = j_{gas}/j_{eq}$.
In particular, at large radii, note that this gas fraction of angular
momentum becomes,
\beq
\left [
\frac{{\dot J}_{gas}}{{\dot J}_{}}
\right ]_{\infty}
=
1- \frac{v_{rA}}{\vinf}
\, .
\label{jgfrac-wdinf}
\eeq
The remaining fraction is carried by the magnetic field,
${\dot J}_{mag}/{\dot J}_{} = v_{rA}/\vinf$.
Since typically $v_{rA}/\vinf \lesssim 1$, the WD67 monopole
field model thus predicts that  most of the angular momentum is lost via the
magnetic field, not the gas.
In their analysis of the solar wind, WD67 obtained an asymptotic
ratio of about 3:1 for angular  momentum of field to gas.

In the somewhat broader context of a monopole field in
a wind with velocity parameterized by a standard `beta' law,
$v(r)/\vinf = (1-R_{\ast}/r)^{\beta}$,
we find for $\beta=1$,
\beq
\left [
\frac{{\dot J}_{gas}}{{\dot J}_{}}
\right ]_{\infty}
=
\frac{R_{\ast}}{R_{A}}
= \frac{1}
{\sqrt{\eta_{\ast} + 1/4} + 1/2}
\, .
\label{fjgasb1}
\eeq
Analogous, but more complicated expressions can be derived for other
values of the index $\beta$.
At the Alfv\'{e}n radius, application of L'Hopital's rule in
eqn.\ (\ref{jgfrac-wd})
gives for general $\beta$,
\beq
\left [
\frac{{\dot J}_{gas}}{{\dot J}_{tot}}
\right ]_{{A}}
= \frac{\beta}
{\beta + 2(R_{A}/R_{\ast} - 1)}
\, .
\label{fjgasra}
\eeq
For example, note that for strong fields in the $\beta=1$ case, the gas fraction
of angular momentum at $R_{A}$ is just half the asymptotic value.

\section{Angular Momentum Loss for a Rotation-Aligned Dipole}

\subsection{MHD simulation parameter study}

While the above simple monopole model is convenient for analytic
study, actual magnetic fields on the sun and other stars can be
far more complex, often represented by many higher order multipoles.
As a first step in extending the above spindown analysis to a more
physically realistic magnetic configuration, let us consider
now the case of a {\em dipole} field with axis {\em aligned}
with that of the stellar rotation.
Relative to a monopole, such an aligned dipole implies variations
in a second spatial dimension, namely co-latitude $\theta$,
as well as radius $r$, but still  retains the axisymmetry that allows
neglect of variations in azimuth $\phi$.
Nonetheless, as shown in papers 1 and 2, the
competition between wind outflow and closed magnetic loops now leads generally
to an inherently complex, time-dependent behavior that is not amenable
to direct analytic study, but instead requires numerical simulation
through the solution of the equations of magnetohydrodynamics (MHD).

The MHD simulations in papers 1 and 2 have examined specifically
the effect of dipole fields in hot, luminous, massive stars with radiatively
driven stellar winds.
Paper 2 presented a detailed parameter study of the competition
among wind, field, and rotation as a function of two dimensionless
parameters, namely the wind magnetic confinement parameter $\eta_{\ast}$
defined in eqn.\ (\ref{esdef}), and a rotation parameter
$W \equiv V_{rot}/V_{orb}$, representing the ratio of the equatorial surface
rotation speed to equatorial orbital speed
$V_{orb} = \sqrt{GM/R_{\ast}}$.
The analysis in paper 2 focussed particularly on the accumulation of
wind material into a dense equatorial disk, confined in nearly rigid
rotation between the Kepler co-rotation radius
$R_{K} \equiv W^{-2/3} R_{\ast}$, and the Alfv\'{e}n radius $R_{A}$.

The remainder of this paper now uses this same parameter study to analyze the angular
momentum loss in this case of a radiation-driven wind from massive star with a
rotation-aligned dipole field at the stellar surface.
The reader is referred to paper 2 for full details of the numerical
method, spatial grid, and assumed stellar and wind parameters.
However, for the convenience of the reader we briefly summarize these below.

For all our calculations we use the ZEUS-3D \citep{StoNor1992} numerical MHD code.
Our implementation here adopts spherical polar
coordinates with 300 grid points in radius $r$ spaced logarithmically and
100 grid points in colatitude $\theta$ with higher
concentration of points near the magnetic equator. We also
assume symmetry in the azimuth $\phi$ direction. To
maintain this 2.5D axisymmetry, we assume the stellar magnetic
field to be a pure dipole with polar axis aligned with the rotation
axis of the star.

As in paper 1 and 2, we consider only the radial component
of radiative force with assumed flow strictly isothermal.
To avoid the effects of oblateness and gravity darkening,
we limit ourselves to  moderate rotation rates applied
to a model with stellar parameters characteristic of
$\zeta$~Pup (see table 1 in paper 1).

\subsection{${\dot J}$ for dipole scaling of Alfv\'{e}n radius}

A key result of paper 2 (see eqn.\ 9 therein) was that in this case of a
dipole field, the equatorial Alfv\'{e}n radius follows the approximate scaling,
\beq
\frac{R_{A}}{R_{\ast}} \approx 0.29 + (\eta_{\ast} + 0.25)^{1/4}
\, ,
\label{radip}
\eeq
which represents an approximate solution of the quartic equation that
arises from requiring $\eta(R_{A}) \equiv 1$ in a dipole ($q=3$) model
with a $\beta=1$ velocity law [cf. eqn.\ (\ref{eta-beta})].

Note in particular that, in the strong confinement limit $\eta_{\ast} \gg 1$,
the Alfv\'{e}n radius in this dipole case now has the scaling
$R_{A}/R_{\ast} \approx \eta_{\ast}^{1/4}$,
instead of the scaling, $R_{A}/R_{\ast} \approx \eta_{\ast}^{1/2}$,
of the monopole model.
As we show below, this modified scaling of the Alfv\'{e}n radius in a
dipole model  has important implications for the associated scaling
of angular momentum loss.

To proceed, let us introduce a basic {\em ansatz} that the overall angular
momentum loss of this rotation-aligned {\em dipole} case can still be
described in terms of the simple WD67 expression of eqn.\ ({\ref{Jtot-wd}),
{\em if} one just uses the dipole-modified scaling (\ref{radip}) for the
Alfv\'{e}n radius.
Specifically, let us define a ``dipole-WD'' (for ``dipole Weber-Davis'')
angular momentum loss as
\beq
{\dot J}_{dWD}
= {2 \over 3}\, \Mdot \Omega R_A^2 \
=  {2 \over 3}\, \Mdot \Omega R_{\ast}^{2}
\left [ 0.29 + (\eta_{\ast} + 0.25)^{1/4} \right ]^{2}
\, .
\label{Jtot-dwd}
\eeq
Here the mass loss rate ${\dot M}$ and wind terminal speed
$\vinf$ used to compute the magnetic confinement parameter $\eta_{\ast}$
are those the star {\em would} have {\em without} a magnetic field,
e.g., as set by the physics of radiative driving.
Even without a field, there is, however, a modest dependence of the
mass loss rate on rotation, found here from numerical simulations of
non-magnetic cases (see figure~8 of paper~2)
to give about a 10\% increase in going from the $W=1/4$ to the $W=1/2$
rotation case.
With these mass loss rates, we use this simple analytic form (\ref{Jtot-dwd})
to scale the numerical MHD simulation results presented below.

\begin{figure*}
\includegraphics[scale=.5]{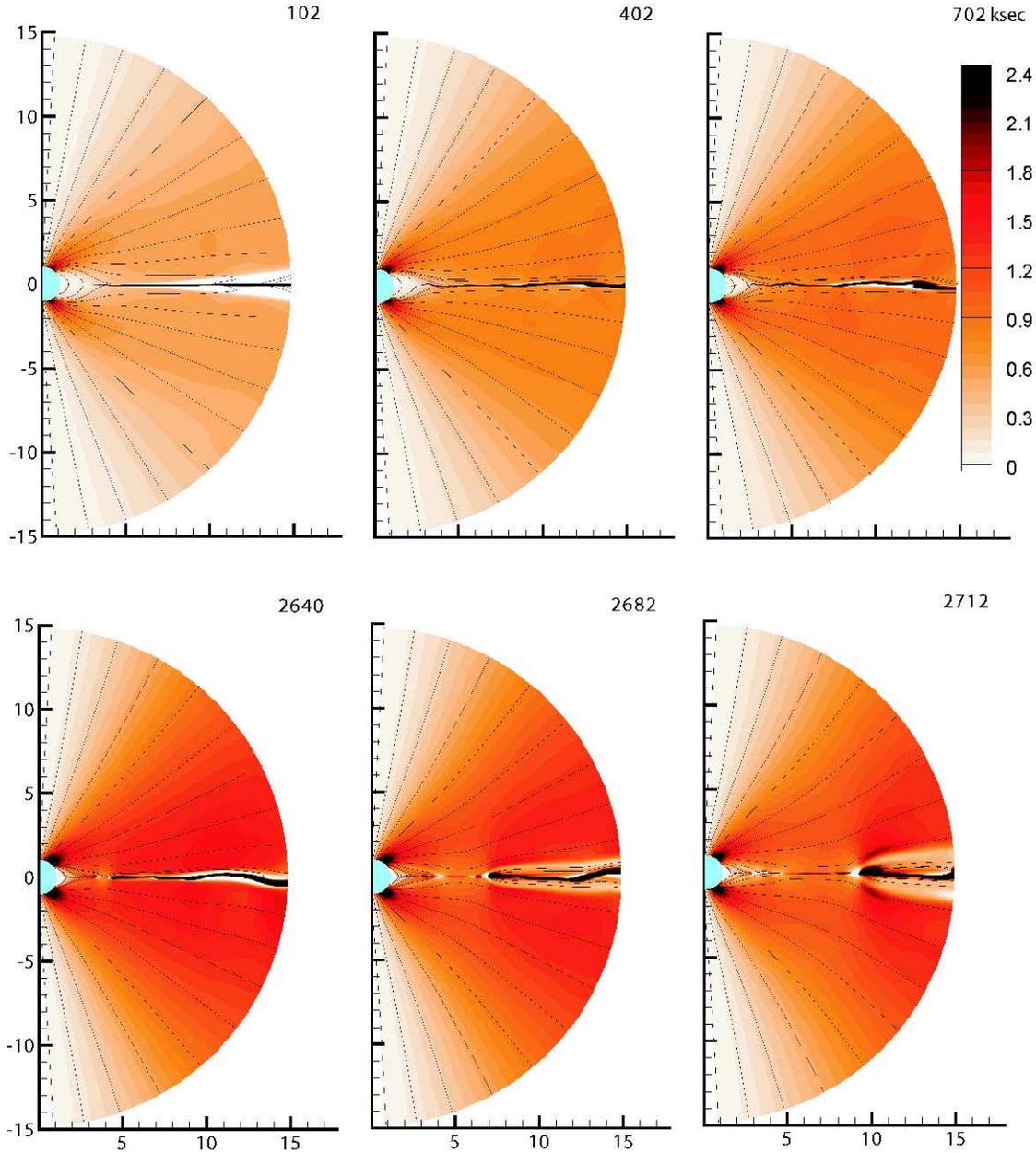}
\caption {
For the
standard case of $\eta_\ast=100$ and $W=0.5$ at the same time
snapshots as in figures~2 and 3 of paper~2,
plots of the spatial variation of $d{\dot J}/d\mu$,
with the colorbar normalized in units of
the predicted  dipole-WD scaling of eqn.\ (\ref{Jtot-dwd})
with an estimated Alfv\`en radius $R_A=3.45 \Rstar$.
\label{dJdmu-eta100}}
\end{figure*}

\subsection{Standard model case: $\eta_{\ast}=100$ and $W=1/2$}

As in paper 2, let us focus first on a standard case with moderately
strong magnetic confinement, $\eta_{\ast} = 100$, and with rotation at
half the critical rate, $W=1/2$.
At any time snapshot of the time-dependent simulation, we can
use eqns.\ (\ref{dJgdmu}) and (\ref{dJmdmu}) to compute,
at each colatitude $\theta$ and radius $r$,
the latitudinal distribution of angular momentum loss associated with the gas
and field,
with their sum thus giving the associated total loss,  $d{\dot J}/d\mu$.

\subsubsection{Spatial distribution and time variation of $d{\dot J}/d\mu$}

For each of the same 6 time snapshots shown in figures~2 and 3 of paper~2,
figure~\ref{dJdmu-eta100} here presents  plots of $d{\dot J}/d\mu$,
with the colorbar normalized in units of
the predicted  dipole-WD scaling of eqn.\ (\ref{Jtot-dwd}).
The changes among the panels emphasize the intrinsic time variability
of the model, with intervals of nearly stationary confinement (upper row)
punctuated by episodes of sudden magnetic breakout (lower row).

Nonetheless, particularly during the confinement intervals,
there is a clear characteristic pattern for the overall distribution
of angular momentum loss.
Near the surface,
$d{\dot J}/d\mu$ is essentially zero
within the close-field equatorial loops,
but this is compensated by a concentration of angular momentum loss
in the open-field regions at mid-latitudes that, as we show below, is
contributed mainly by the magnetic component.
Moreover, further from the star, there is a broad latitudinal distribution from the
magnetic component combined with an equatorial concentration from the gas
component.

\begin{figure*}
   \includegraphics[scale=0.5]{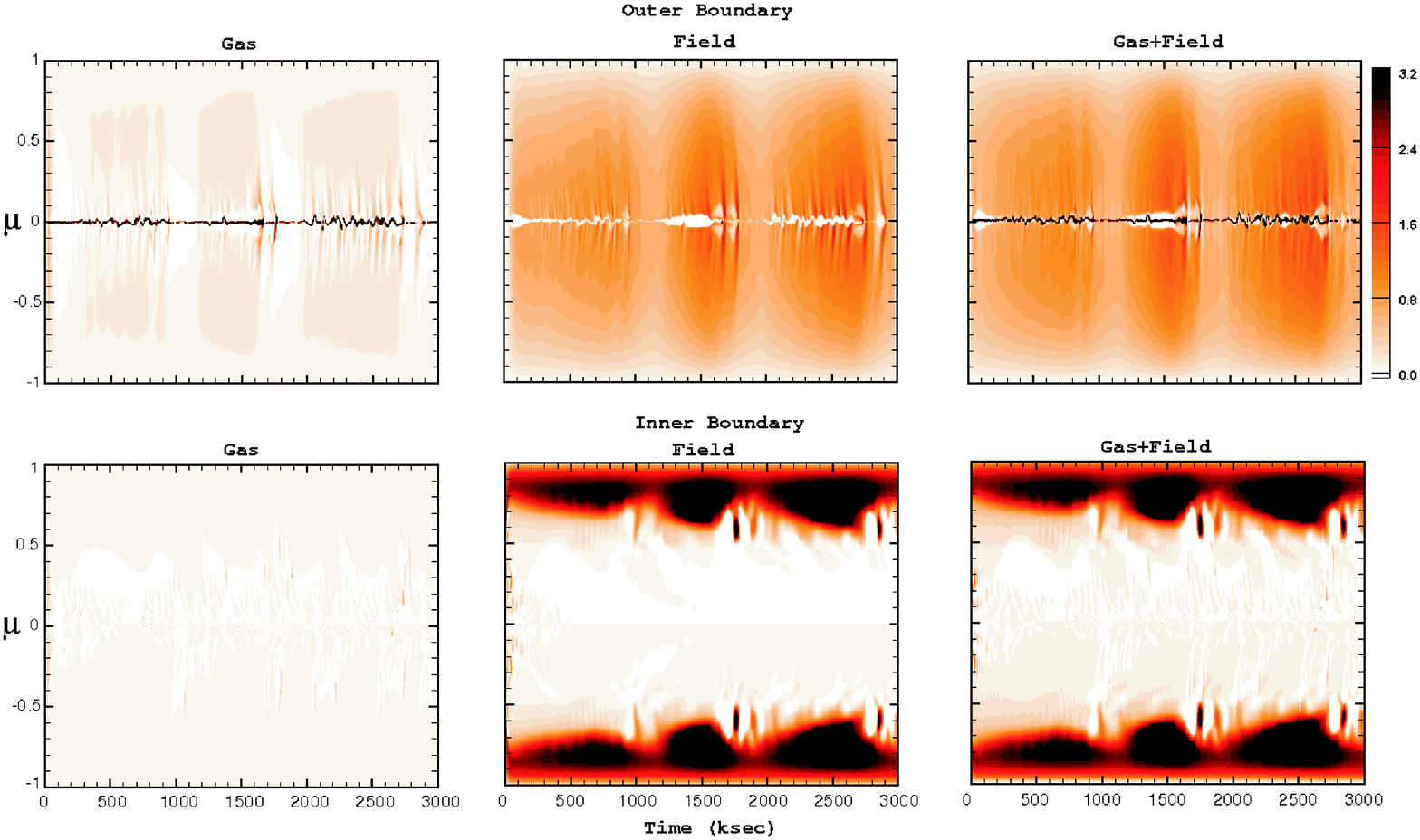}
\caption {
Plot of the latitude and
time dependence of the gas and field components of $d{\dot J}/d\mu$ through
the inner boundary ($r = R_{\ast}$; bottom panels) and
the outer boundary ($r= 15 R_{\ast}$; top panels),
for the standard model case. The colorbar is again
normalized by the dipole-WD scaling of eqn.\ (\ref{Jtot-dwd}).
\label{djdmu-bndy-eta100}
}
\end{figure*}


The colorscale in figure~\ref{djdmu-bndy-eta100} illustrates this
latitudinal distribution
and time variation of the gas, magnetic and total (gas + magnetic)
angular momentum loss $d{\dot J}/d\mu$ along both the inner
and outer boundary
for this standard model case.
For the inner boundary, the broad white region at low latitudes emphasizes
quite clearly now
that the closed magnetic loops above the equatorial surface represent
a kind of ``dead zone'' with little or no angular momentum loss.
Instead, the mid-to-high latitudes of open field carry a strong
concentration of the surface angular momentum loss.
In contrast, at the outer boundary, there is a broad latitudinal distribution of
angular momentum loss from the field, together with a equatorial
concentration from the gas.
 Both the inner and outer boundary distribution of angular momentum also show a
clear time variation associated with the ca. 1 Msec cycle of confinement,
buildup, and release of material trapped in closed magnetic loops.

\subsubsection{Time variability of latitudinally integrated ${\dot J}$}

\begin{figure*}
  \includegraphics[scale=0.7]{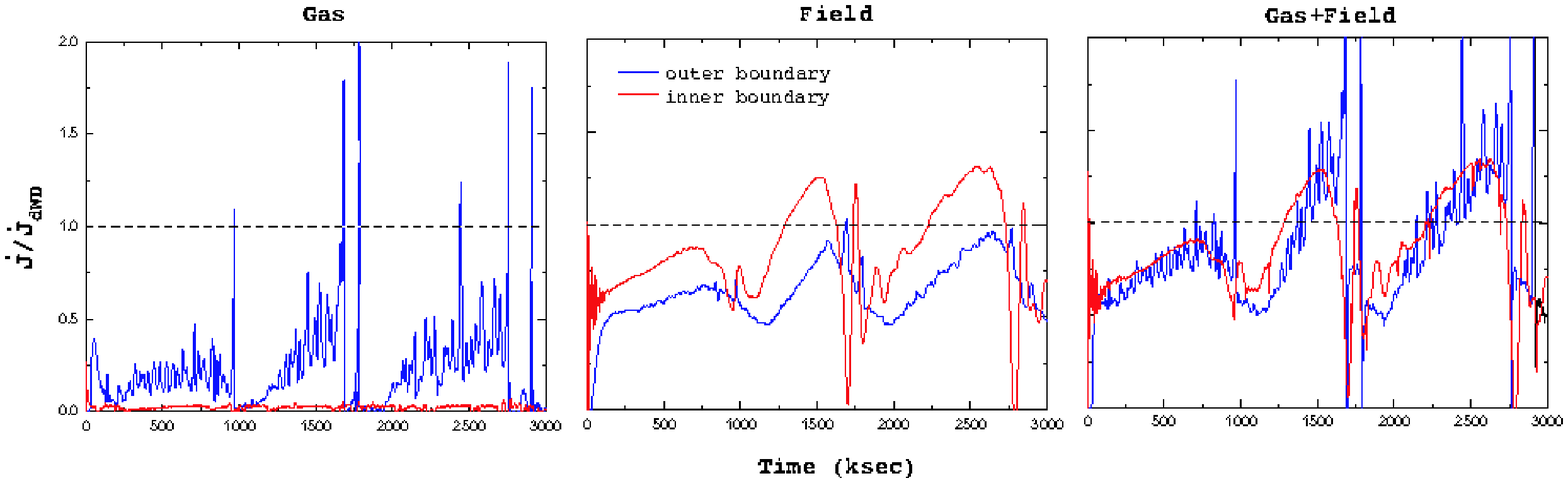}
\caption {
Time dependence of latitudinally integrated angular momentum loss
${\dot J}$ through both
the inner
and outer boundary,
computed for the gas (left panel), field (middle) and the
total for gas + field (right), in the standard model case.
\label{jdot-bndy-eta100}
}
\end{figure*}

The line plots in
figure~\ref{jdot-bndy-eta100} compare the time dependence of the
latitudinally integrated angular momentum loss, ${\dot J}$, for the
gas, field, and total components, again evaluated at the inner
(red curves)
and the outer boundaries
(blue curves).
At the inner boundary, the gas component is negligible, with most of the
angular momentum associated with the field.
At the outer boundary, the gas component is highly variable, but still
generally small compared to the field.
The total angular momentum loss shows a nearly periodic variation,
characterized by a gradual ramping up that ends in a sudden drop,
with however a slight relative lag in the outer vs. inner variation.
This lag reflects a cycle of storage and release of
angular momentum within both the circumstellar field and gas.

The colorscale plots in figure~\ref{jdot-r-vs-t} show also the full
radius and time variation for the gas, magnetic,  and total angular
momentum loss, with the colorbar again normalized in units of the
predicted dipole-WD scaling of eqn.\ (\ref{Jtot-dwd}).
The results show quite vividly the intrinsic time variability,
particularly for the gas component, which varies from intervals of
little or no angular momentum loss, to a series of radially ejected
streams, punctuated by a strong interval of loss during the magnetic
breakout.
The generally pale color reflects the fact that the overall level of
gas angular momentum loss is a small fraction of the total expected
from the dipole-WD scaling, particularly near the stellar surface.
By comparison, the magnetic component is stronger and less variable.
Overall, we see that the total angular momentum loss varies by about
50\% above and below the predicted dipole-WD value from eqn.\
(\ref{Jtot-dwd}).

\begin{figure*}
\includegraphics[scale=0.53]{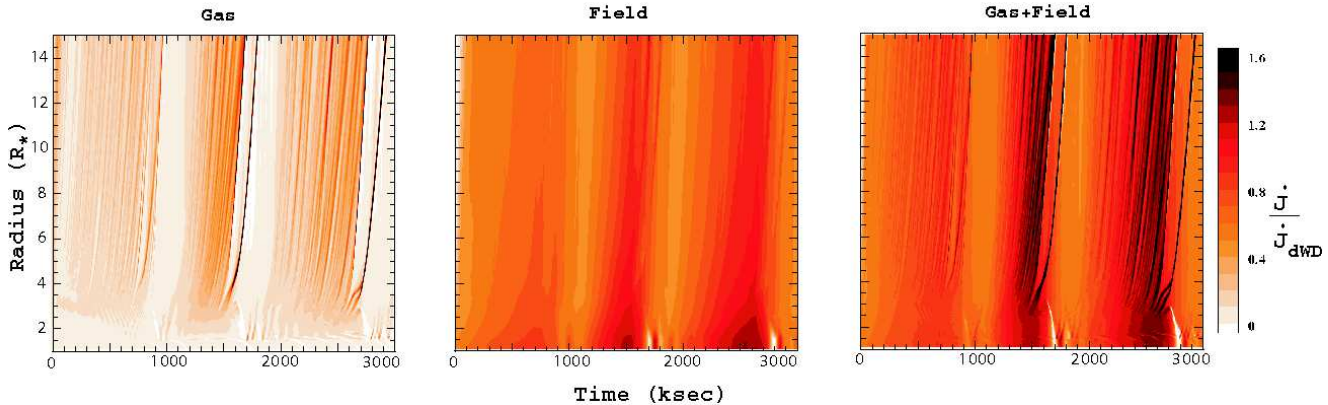}
\caption {
plots of the full radius and time variation of
${\dot J}$, again computed for gas, magnetic, and total components in
the standard model,  with the colorbar normalized in units of
the predicted  dipole-WD scaling of eqn.\ (\ref{Jtot-dwd}).
\label{jdot-r-vs-t}
}
\end{figure*}

\subsubsection{Spatial variation of time-averaged ${\dot J}$
for gas vs. magnetic component}

\begin{figure}
\includegraphics[scale=1.0]{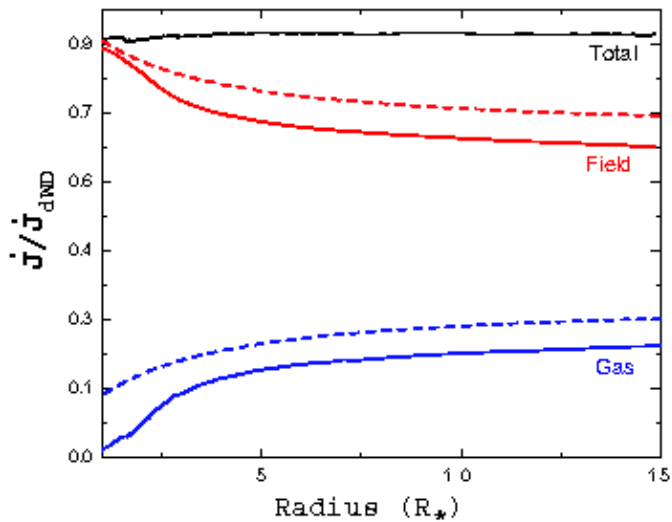}
\caption {Radial variation of gas, field, and total
angular momentum loss, again scaled by the total loss in the dipole-WD model.
The solid curves show the time-averages from the standard numerical
simulation model, while the dashed curves compare the
corresponding WD scalings implied by eqn.~(\ref{jgfrac-wd}),
using the Alfv\'{e}n radius $R_{A}$ from the dipole eqn.~(\ref{radip})
assuming a $\beta=1$ velocity law and $\eta_{\ast} = 100$.
\label{jdot-time-avg}
}
\end{figure}

To show more clearly the overall spatial variation,
figure~\ref{jdot-time-avg} plots the  radial
dependence of the time-averaged angular momentum loss for
both the gas (blue curve) and magnetic field (red curve),
along with the total loss (black curve).
Note that this time-averaged total
loss is nearly constant in radius,
at a value that is remarkably
close -- about 90\% --
to the simple analytic dipole-WD predicted scaling!

Moreover, much as found in the WD67 monopole model, the angular
momentum loss by the gas increases with radius, but still
remains everywhere relatively small compared to the magnetic component.
Indeed, the dashed curves compare the
corresponding WD scalings given by eqn.~(\ref{jgfrac-wd}), using the
Alfv\'{e}n radius $R_{A}$ from the dipole eqn.~(\ref{radip}),
and assuming a $\beta=1$ velocity law and $\eta_{\ast} = 100$.

The good general agreement shows that,
despite the complex time variation of the dipole case, the overall,
time-averaged scaling of angular momentum loss can be quite well
modeled through the simple monopole scalings of WD, as long as one
just accounts for the different scaling of the Alfv\'{e}n radius with
the magnetic confinement parameter $\eta_{\ast}$.

\subsection{Dependence on magnetic confinement and rotation parameters,
$\eta_{\ast}$ and $W$}

To build on this success in using the dipole WD scaling to characterize
angular momentum loss in this standard case of moderately strong
confinement and rotation,
let us now examine how well this simple scaling agrees with the
numerical simulation results for variations of magnetic
confinement parameter $\eta_{\ast}$ and rotation parameter $W$.

\begin{figure}
\includegraphics[scale=1.0]{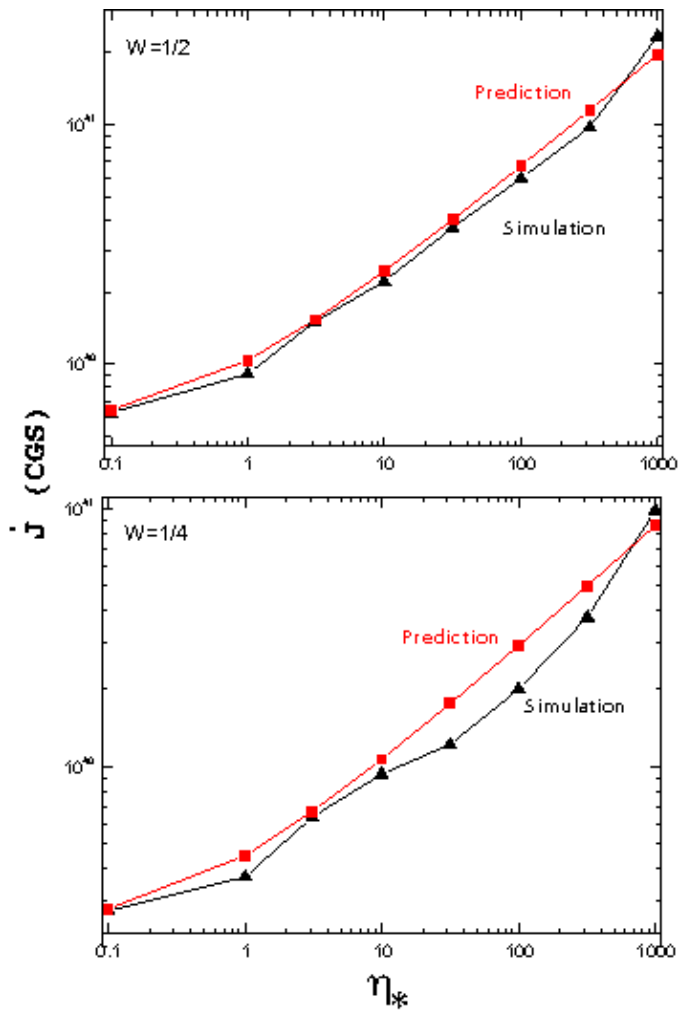}
\caption {Time-averaged angular momentum loss
for all the models (triangles), plotted vs. $\eta_{\ast}$ (on a log-log scale), for
both the $W=1/4$ (left) and $W=1/2$ rotation cases.
The squares compare the scalings predicted by the dipole-WD
approximation (\ref{Jtot-dwd}).
\label{Jdot-tot-vs-eta}}
\end{figure}

For the rotation cases $W=1/4$ and $W=1/2$, the lower and upper panels
of figure~\ref{Jdot-tot-vs-eta} compare the variation of total, time-averaged
angular momentum loss rate vs.
$\eta_{\ast}$ (on a log-log scale)
for both the numerical
simulations (triangles) and analytic form (\ref{Jtot-dwd}) (squares).
The overall agreement is remarkably good for both rotation cases,
confirming that, quite independent of the rotation parameter $W$,
this very simplified form (\ref{Jtot-dwd}) provides a good description of
the scaling of the average angular momentum loss in this case of aligned
dipoles.

Note that the ordinate axes in figure~\ref{Jdot-tot-vs-eta} are labeled
in CGS units computed for the specific stellar model used in the
simulations.
But these specific values are essentially arbitrary.
For any star of interest, the appropriate physical values can be
readily derived from the dipole-WD scalings in eqn.~(\ref{Jtot-dwd}).
Indeed, the plot can likewise be characterized as giving the inverse
of the spin-down time, which in the dipole-WD model has the specific scaling
\beq
\frac
{\tau_{spin}}
{\tau_{mass}}
\approx
\frac
{\frac{3}{2} k}
{\left [ 0.29 + (\eta_{\ast} + 0.25)^{1/4} \right ]^{2}}
\, .
\label{tauspin-dwd}
\eeq

Figure~\ref{Jdot-fraction-vs-eta} shows the fraction of the gas and magnetic
components of angular momentum loss at the outer boundary for each
model,  plotted vs. magnetic confinement parameter $\eta_{\ast}$
(again on a log scale),
for both the $W=1/4$ (left) and $W=1/2$ (middle) rotation cases.
For comparison, the right panel plots the corresponding large-radius
scalings for the pure monopole model with a $\beta=1$ velocity
law, as given by eqn.~(\ref{fjgasb1}).
For low and moderate magnetic confinement, $\eta_{\ast} \lesssim 30$,
there is good general agreement between the analytic scalings and the
numerical simulation results;
but for stronger confinement, the
numerical results show the gas fraction reaching a minimum and then {\em
increasing} with increasing $\eta_{\ast}$.

The reasons for this increase are not entirely clear, but likely could
be the result of the channeling and confinement of wind gas into
the equatorial, nearly rigid-body disk discussed in paper 2.
The magnetic torquing that spins this material up into a rigid disk
represents a transfer of angular momentum from field to gas that has
no parallel in the pure-outflow, monopole models of WD.
Once sufficient material accumulates in this disk, the outward
centrifugal force overwhelms the inward confinement of magnetic
tension, leading to a breakout of this material that now carries outward
a strong gas component of angular momentum loss.

\begin{figure*}
\includegraphics[scale=0.7]{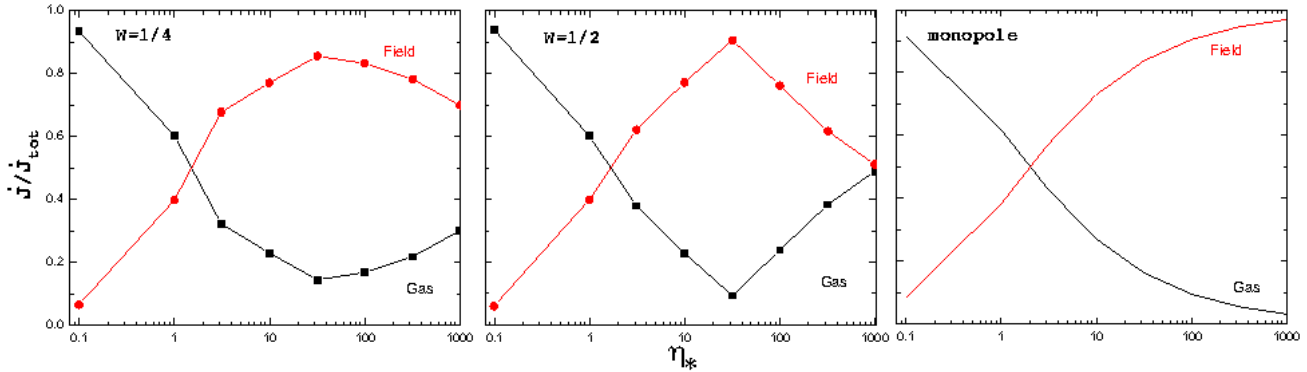}
\caption {Fraction of the gas and magnetic components of angular
momentum loss, plotted vs. magnetic confinement parameter $\eta_{\ast}$
(on a log scale), for both the $W=1/4$ (left) and $W=1/2$ (middle) rotation
cases. The right panel shows analytic scalings for the
pure monopole model with a $\beta=1$ velocity
law, as given by eqn.~(\ref{fjgasb1}).
\label{Jdot-fraction-vs-eta}
}
\end{figure*}

\section{Discussion}

\subsection{Role of mass-loss ``Dead Zone''}

A principal result of the above parameter study is that the overall
level of angular momentum loss from an early-type star with a rotation-aligned
dipole can be  well described by the simple dipole-modified WD scaling given in
eqn.~(\ref{Jtot-dwd}).
In this formulation, the mass loss rate and wind terminal speed
used to compute the magnetic confinement parameter $\eta_{\ast}$
and associated spindown are those the star {\em would} have
{\em without} a magnetic field, as set by the physics of radiative driving.

As shown in figure~8 of paper~2, the {\em actual} net mass loss rates
in this parameter study of dipole winds show a significant {\em decline}
with increasing confinement parameter $\eta_{\ast}$, fit roughly by the
scaling relation [given in eqn.\ (24) of paper~2],
\beq
\frac{{\dot M_{\rm {B}}}}{{\dot M_{\rm{B}=0}}}
\approx 1 - \sqrt{1-\Rstar/R_{\rm{c}}}
+ 1 - \sqrt{1-0.5*\Rstar/ R_{\rm{K}}}
\label{mdred}
\, ,
\eeq
where
$R_{\rm{c}} \approx R_{\ast} + 0.7 (R_{\rm{A}} - R_{\ast})$ is a
maximum ``closure'' radius of magnetic loops, and
$R_{\rm{K}} = R_{\ast}/W^{2/3}$ is the Kepler co-rotation radius.
The former accounts for the effect of the mass loss ``dead zone''
of closed magnetic loops, while the latter corrects for the eventual
centrifugal breakout that can occur from some initially closed loops
above the Kepler radius.

In previous discussions of rotational spindown of magnetic winds,
this dead zone has generally been presumed \citep[e.g.,][]{Mes1968a,Don2006b}
to lead to a downward modification in the net angular momentum
loss that would otherwise occur, based on the notion that the mass trapped
in these closed loops does not (at least for loops closing below the Kepler
radius) escape from the star,  and  thus should not contribute to the angular
momentum loss.

This notion seems partly based on the perception that the gas
itself is the principal direct carrier of the angular momentum loss.
But both the WD67 analysis and the simulations here show
that the dominant effect of the gas is indirect, inducing an azimuthal field
component that then carries the bulk of the angular momentum loss,
particularly near the stellar surface.
Figure~\ref{djdmu-bndy-eta100} shows that, even for this magnetic
component,  the closed loops at low latitudes do indeed represent a
dead zone for loss of angular momentum, as well as mass.
The net effect, however, seems merely to shunt a fixed total amount of
angular momentum towards the mid-latitudes, carried by the azimuthal
twisting of the open magnetic field.
As the wind expansion opens up the field beyond the Alfv\'{e}n radius,
this transport of angular momentum spreads to cover all latitudes
(cf. right vs. left panels of figure~\ref{djdmu-bndy-eta100}), and
includes an increasing (but generally still minor) component for the
gas (see figures~\ref{jdot-bndy-eta100},
\ref{jdot-r-vs-t}, \ref{jdot-time-avg}, and
\ref{Jdot-fraction-vs-eta}).

So an important lesson of the above parameter study is that this
additional dead-zone reduction in the net ${\dot J}$ does {\em not}
apply to the dipole-modified WD scaling form (\ref{Jtot-dwd}).
In a sense, it is already incorporated in the reduction associated with
the change from the monopole scaling
${\dot J} \sim {\dot M} \eta_{\ast}$
to the dipole scaling
${\dot J} \sim {\dot M} \sqrt{\eta_{\ast}}$.
To see this, note that, if we ignore the minor rotational correction
of the Kepler term, the net mass loss reduction given by
eqn.~(\ref{mdred}) has the
strong-confinement ($\eta_{\ast} \gg 1$) scaling
\beq
\frac{{\dot M_{\rm {B}}}}{{\dot M_{\rm{B}=0}}}
\approx 1 -
\left ( 1- \frac{\Rstar}{2 R_{\rm{c}}} \right )
\approx \frac{\Rstar}{1.4 \,R_{\rm{A}}}
~~ ; ~~ R_{\rm{A}} \gg \Rstar
\, .
\label{mdred-egg1}
\eeq
If we now use this to apply a ``dipole dead-zone'' correction to
the standard {\em monopole} scaling for angular momentum loss, we
then find
\beq
{\dot J}
\sim {\dot M_{\rm {B}}}   R_{A}^{2}
\sim {\dot M_{\rm {B}=0}} R_{A}
\sim {\dot M_{\rm {B}=0}} \sqrt{\eta_{\ast}}
\, ,
\eeq
which thus reproduces the dipole scaling using the {\em non}-magnetic
value for the mass loss rate!

The upshot then is that the dipole-modified scaling
(\ref{Jtot-dwd}) using the non-magnetic mass loss rate
effectively already accounts for the dead-zone reduction of the
actual mass loss.

\begin{table*}
\caption{Estimated Spindown Time for Selected Known Magnetic Stars.}
\begin{minipage}{\linewidth}
\renewcommand{\thefootnote}{\thempfootnote}

\begin{center}
\begin{tabular*}{1.\textwidth}{@{\extracolsep{\fill}}|c||c|c|c|c|c|c|c|c|c|}
\hline
Star \footnote{References: $^1$\citet{Don2002}; $^2$\citet{Don2006b};
$^3$\citet{Nei2003} and \citet{SmiBoh2007}; $^4$\citet{KrtKub2006}; $^5$\citet{Kho2007}}
& $M/M_{\odot}$ & $R_{\ast}/R_{\odot}$ &P (days)& k & ${\dot M} \, (10^{-9} M_{\odot}/yr)$
  &$\vinf~(1000~km/s)$ &  $B_{p}~(kG$) & $\eta_{\ast}$   & { $\tau_{spin}$(Myr)} \\
  \hline
$\theta^{1}$~Ori~C $^1$& 40 & 8&15.4 & 0.28 & 400 & 2.5 & 1.1 & 15.7& {\bf 8} \\
 \text{HD191612} $^2$& 40 & 18 &538& 0.17 & 6100 & 2.5 & 1.6 & 7.6 & {\bf 0.4} \\
$\zeta$~Cas $^3$& 8 & 5.9 &5.37& 0.1 & 0.3 & 0.8 & 0.34 & 3200 & {\bf 65.2} \\
$\sigma$~Ori~E $^4$& 8.9 & 5.3 &1.2& 0.1 & 2.4 & 1.46 & 9.6 & 1.4 $\times 10^5$ & {\bf 1.4}\\
$\rho$ Leo $^5$&22 & 35&7-47&0.12&630&1.1&0.24&20&\bf{1.1}\\
\hline
\end{tabular*}
\end{center}
\end{minipage}
\end{table*}

\subsection{Spindown time}

From the above, the overall stellar spindown time is predicted to
follow the scaling in eqn.~(\ref{tauspin-dwd}).
In the strong-confinement limit, this gives

\beqa
\nonumber
{\tau_{spin}}
&\approx&
{\tau_{mass}}
\frac
{\frac{3}{2} k}
{\sqrt{\eta_{\ast}}}
\\
\nonumber
&\approx&
\frac
{\frac{3}{2} k M}
{\Beq R_{\ast}}
\, \sqrt{\frac{\vinf}{{\dot M}}}
\\
&\approx&
1.1 \times 10^{8} {\rm yr}
~ \frac {k}{B_{p}/kG}
\, \frac{M/R_{\ast}}{M_{\odot}/R_{\odot}}
\, \sqrt{\frac{V_{8}}{{\dot M}_{-9}}}
\, .
\label{tauspin-dwd-egg1}
\eeqa
Comparison with the monopole scaling (\ref{tauspin-wd}) shows that the
spindown is no longer independent of the mass loss rate, but now
varies with its inverse square root.
In addition, the dependencies on surface field and radius is now inverse
linear instead of inverse square.

In practice, application of this scaling relation requires
observational and/or theoretical inference of the various parameters.
The last equality in (\ref{tauspin-dwd-egg1}) gives characteristic
B2-star scalings  for the mass loss,
${\dot M}_{-9} \equiv {\dot M}/(10^{-9} \, M_{\odot}$/yr),
and wind speed, $v_{8} = \vinf/(10^{8} \,$cm/s).
Note that the magnetic field is now quoted as a surface value at the
{\em pole}, $B_{p}$, and is scaled in kG.
This is a typical value for known magnetic massive stars, as inferred from
Stokes V measurement  of photospheric lines with circular polarization
by the Zeeman effect \citep[see, e.g.,][]{Don2002}.

The mass to radius ratio $M/R_{\ast}$  can best be estimated from
atmosphere models for the given spectral type, but generally
for massive main sequence stars this
should be just somewhat above the solar ratio.
The moment of inertia constant $k$ can be estimated from stellar
structure models, and should be typically
$k \approx 0.1$, perhaps somewhat higher near the zero-age main
sequence (ZAMS), and then decreasing by up to 50\% with
age \citep{Cla2004}.\footnote{
Note that this presumes solid-body rotation; if the stellar
envelope should decouple from the core, the effective k could be much
lower (to account for the lower envelope mass).
But a strong internal magnetic field should generally be effective in
enforcing near-rigid rotation in the interior.}

Perhaps the most difficult parameters to infer are those for the
stellar wind.
Fortunately, these enter only in proportion to the
square root of the ratio of wind speed to mass loss rate.
But because the {\em actual} values for both of these are likely to be
strongly affected by the magnetic field, it may generally be better not to
infer them from direct observations for the actual star in question, but
rather to use the inferred spectral type to apply observational or
theoretical scaling laws for the values in similar {\em non}-magnetic stars.

Table~1 lists  spindown times based on estimated parameters for a
sample of known magnetic hot stars.
The first two known magnetic O-stars, $\theta^{1}$~Ori~C
\citep{Don2002} and
HD~191612 \citep{Don2006b},
are both slow rotators, with periods of respectively 15~d and 538~d.
$\theta^{1}$~Ori~C is thought to be quite young,
less than 0.2~Myr, and so still on the ZAMS,
while HD~191612 is thought to be more evolved, with an age of 2-3~Myr.
Using parameters quoted in \citealt{Don2006b}, we infer
corresponding spindown times of
respectively $\sim$8~Myr and $\sim$0.4~Myr.
This implies that magnetic wind braking seems unlikely to explain the
slow rotation of  $\theta^{1}$~Ori~C, but it does seem potentially
relevant for HD~191612.
Alternatively, magnetic effects during star formation could have lead to
an initial slow rotation.

For He-strong stars, a key object is the B2V star $\sigma$~Ori~E.
This has an estimated polar field strength of $B_{kG} \approx 9.6$,
and with remaining parameters
as in table~1,
we estimate a spindown time of $\sim$1.4~Myr.
As a main sequence B2 star, its age is likely
comparable to this.
The rotation period, 1.2~d, is still quite short, about twice that
for critical rotation, implying only a moderate net spindown since formation.

\subsection{Extension to non-aligned dipoles and higher
multipoles}

However note that, as is typical of magnetic massive stars, most of the
above cases exhibit dipole fields with a significant tilt angle to the
rotation axis.
MHD simulation of such a tilted dipole requires accounting for 3-D,
non-axisymmetric outflow as the polar field sweeps around in
azimuth. This remains a challenge for future studies.
Without such simulations, we can only offer some general speculations
on how such a tilt might affect the spindown.
Generally, it seems that angular momentum loss should be modestly enhanced,
because of the factor two stronger polar field.

But another factor might be the {\em open} nature of this polar field.
One might expect this to lead to a larger magnetic moment arm,
perhaps even following the stronger monopole scaling,
${\dot J} \sim \eta_{\ast}$,
rather than the dipole form ${\dot J} \sim \sqrt{\eta_{\ast}}$.
But the analysis in the Appendix suggests that such magnetic
polar-axis fields in an oblique rotation case should actually
follow a {\em weaker} spindown scaling,
${\dot J} \sim \eta_{\ast}^{1/3}$.
This implies that, much as in the aligned-dipole case,
the overall,  spherically averaged loss rate for an
oblique-dipole wind should still be dominated by the regions surrounding
closed loops,  with a net scaling that thus is similar to the
aligned case.

In some magnetic hot stars, the inferred field is manifestly
non-dipole.
For example, the B2IVp He-strong star HD~37776 (V901 Ori) has been
inferred to have quadrupole field that dominates any dipole component,
with peak strength of $\sim$10~kG
\citep{ThoLan1985}.
The luminosity class IV would normally imply an evolved, main-sequence
star, but the still moderately short rotation period of 1.5387~d
seems to suggest little spindown.
If we assume a somewhat larger radius
and higher mass loss rate than $\sigma$~Ori~E, so that the
non-magnetic parameters in (\ref{tauspin-dwd-egg1}) may be near unity,
then applying the inferred field of 10~kG gives a spindown
time of $\sim$1.1~Myr.

Moreover, although detailed predictions must await 3-D MHD simulations of
such a quadrupole (or higher multipole) case, one might infer
that the steeper radial decline of the quadrupole field
should lead to a weaker spindown.
For a general field scaling with $r^{-q}$, with $q=3$ for a dipole and
$q=4$ for a quadrupole, the ratio of magnetic to wind energy should
decline as $r^{2-2q}$, implying an Alfv\'{e}n radius that scales as
$R_{A} \sim \eta_{\ast}^{1/(2q-2)}$, or $R_{A} \sim \eta_{\ast}^{1/6}$
for a quadrupole.
In the strong-confinement limit, the expected spindown scaling should
thus become
\beq
\frac
{\tau_{spin,quad}}
{\tau_{mass}}
\approx
\frac
{\frac{3}{2} k}
{\eta_{\ast}^{1/3}}
\, .
\label{tauspin-quad}
\eeq
Compared to a dipole of the same surface strength,
the spindown time for a quadrupole would thus be about a factor
$\eta_{\ast}^{1/6}$ longer.
This might seem like a weak correction, but
for He-strong stars of spectral type B2, the low mass loss means that
a surface field of 10~kG gives a confinement parameter of
$\eta_{\ast} \approx 10^{7}$,
thus implying a factor $\sim$10 times longer spindown for a quadrupole
vs. a dipole of same surface strength.
In this context, the expected spindown time for HD~37776 becomes of
order $\sim$10~Myr.

On the other hand, the presence of a more complex field might make it
easier for the wind to break open the magnetic flux associated with
large-scale dipole component, and so allow a more extended magnetic
moment arm.
In principal, this could even lead to a
{\em stronger} spindown effect.
\citet{Mik2008}
have recently reported a direct
measurement of rotational spindown in HD~37776.
They infer a $17.7 \pm 0.7$~second increase in the 1.5387~day period over a
span of 31~years, translating to a spindown time of just 0.23~Myr!
This is substantially shorter than the above dipole  estimate from wind torquing.
Indeed,
the luminosity class IV would normally imply an evolved, main-sequence
star, but the still moderately rapid rotation together with
the very rapid spindown seem to require a very young age.

Overall, the study of angular momentum loss in these more non-aligned
dipole or higher-order multipole cases must await further, 3-D
MHD simulation studies.

\subsection{No sudden spindown during breakout events}

In a few magnetic stars, there appears to be evidence
for sudden change in rotation period.
For the rapidly rotating
Ap star CU~Virginis (HD~124224), \citet{Tri2008}
cite radio observations indicating changes in rotational phase over
20 years, identified with two discrete period increases of 2.18~s in
1984 and about 1~s in 1999.
In terms of the still-rapid rotation period 0.52~days,
the associated average spindown timescale over this
15~year timespan is quite short, only about 300,000~years.
These sudden period changes could be associated with changes
in the star's internal structure \citep{Ste1998};
but \citet{Tri2008}  suggest they might also be the result
of a sudden emptying of mass accumulated in the star's magnetosphere.

The simulations here do indeed show repeated episodes of substantial
emptying, but
careful examination indicates that these do not lead to sudden jumps in the angular
momentum of the star itself.
Both figures \ref{djdmu-bndy-eta100} and \ref{jdot-bndy-eta100} show,
for example, that angular momentum loss
through the stellar surface varies smoothly through the cycle of
magnetosphere build-up and release, and if anything is actually {\em less} during
the sudden breakouts that characterize magnetospheric emptying.
In terms of stellar rotation, the  outward transfer of angular momentum
to the magnetosphere occurs gradually, as the magnetosphere fills up,
effectively increasing the moment of inertia of the star+magnetosphere system.
As such, when the emptying does occur,  it merely represents the final
escape for angular momentum that had already been lost to the star.
Moreover, most of the angular momentum is not even contained in the
trapped gas, but rather in the stressed magnetic field, which
figure 3b shows varies quite smoothly from the stellar surface.

Thus, even apart from questions of the magnitude
of angular momentum loss needed to explain the average spindown inferred
for CU~Vir,  magnetospheric emptying does not seem well-suited to
explaining the claimed suddenness of rotation changes in this star.

\subsection{Comparison with previous spindown analyses}

Past studies of wind magnetic spindown have primarily focussed on
cool, low-mass, solar-type stars.
The main analysis aimed specifically at spindown for hotter, higher-mass  star,
the main analysis was by FM84,
who derived 1-D, steady-state solutions for the
equatorial flow of a CAK-type line-driven wind from a rotating hot
star with a Weber-Davis style {\em monopole} magnetic field.
To ensure their steady solutions passed smoothly through the various
flow critical points (i.e., those associated with Alfv\'{e}n,
slow-mode, and fast-mode MHD waves, as well as the usual CAK critical point),
they had to use a quite elaborate numerical iteration scheme;
as such, they do not quote any simple scaling forms for the Alfv\'{e}n
radius and associated spindown rate.

Instead, their Table 1 lists numerical results for a set of 16 models with
various rotation rates and field strengths, assuming fixed stellar and wind
parameters for a typical O-supergiant.
We find here that these results can be generally well fit (within
ca.\ 10-15\%) by the simple general monopole scaling for $R_{A}$,
eqn.\ (\ref{rawd-beta1}).

However, FM84 quote spindown times as low as 60,000~years, i.e.,
for their case 4, with field strength of 1600~G, and mass loss rate of
$5.32 \times 10^{-6} \msbyr$.
The associated confinement parameter is $\eta_{\ast} \approx 80$.
Since  the angular momentum loss rate for an aligned dipole
is a factor $\sqrt{\eta_{\ast}}$ lower than for a monopole, we see
that a dipole with a similar surface field strength (1600~G) as FM84
model 4 would have about a factor 9 longer spindown time, or now about
0.5~Myr.
In contrast, their weaker-field model, e.g.\ case~1 with only 200~G,
has a confinement parameter $\eta_{\ast} \approx 1.7$, implying only
about a 30\% dipole increase above the 2~Myr spindown time quoted for this
case by the FM84 monopole analysis.
Overall, the dipole modifications in spindown rate found here suggest
that the upper limits on surfaced field strength inferred by \citet{Mac1992}
for rotating hot-stars are likely to be too low.

For cooler, solar-type stars with coronal-type pressure-driven winds,
the literature on wind magnetic spindown is more extensive, and
includes both semi-analytic studies \citep{Oka1974,Mes1984,TouPri1992}
and numerical iterations or simulations, \citep{Sak1985,WasShi1993,KepGoe2000,MatBal2004}.
In addition to spindown by a magnetized coronal wind during the star's
main sequence phase,
there has been extensive study of angular momentum loss during
pre-main-sequence accretion through a magnetized disk during a T-Tauri star
(TTS) phase.
Most recently, \citet{MatPud2008}
report on 2-D MHD simulations of
spindown by an aligned dipole field in a coronal wind during the TTS phase.
With some translation for differences in notation and parameter
definition, many of their results seem quite analogous to those
reported here for hot-star winds.
In particular, their eqn.\ (14) for the Alfv\'{e}n radius is quite
similar to the dipole scaling in eqn.\ (\ref{radip}) here,
just replacing $\vinf$ with the (numerically comparable) surface escape
speed $v_{esc}$, with their numerical best-fit exponent $m=0.223$ very similar
to our $1/4=0.25$.

\section{Summary and Future Outlook}

This paper analyzes the nature of angular momentum loss by radiatively driven
winds from hot-stars with a dipole magnetic field aligned to the stellar
rotation axis.
It applies our previous MHD simulation study from paper 2, which
consisted of a 2-parameter series of models,
dependent on the  critical rotation ratio $W$,
and on the magnetic confinement parameter $\eta_{\ast}$.
Key results can be summarized as follows:
\begin{itemize}
\item
As shown in figure~\ref{Jdot-tot-vs-eta}, for both slow ($W=1/4$) and
moderate rotation ($W=1/2$), the time-averaged, total angular momentum
loss rate, ${\dot J}$,
follows closely a dipole-modified version of the Weber-Davis scaling
of eqn.\ (\ref{Jtot-wd}).
\item
Specifically, in the dipole case,
the Alfv\'{e}n radius has a weaker dependence on magnetic
confinement parameter,
with a strong confinement scaling as $R_{A} \sim \eta_{\ast}^{1/4}$
vs.\ as $R_{A} \sim \sqrt{\eta_{\ast}}$ for the monopole
[cf. eqns.\ (\ref{radip}) and (\ref{rawd-beta1})].
\item
This leads to a dipole scaling for ${\dot J}$, given by eqn.\ (\ref{Jtot-dwd}),
that is weaker than for a monopole, by a factor $1/\sqrt{\eta_{\ast}}$
in the strong-confinement limit.
It also implies a correspondingly longer spindown time.
\item
As in the WD67 monopole case, the total angular momentum loss is
generally dominated by the magnetic component.
However, in the strongest confinement models, there is trend toward
increasing contribution by the gas, apparently a consequence of
the eventual breakout of equatorially trapped material.
\item
The dipole nature of the field gives the angular momentum transport
a complex variation in radius, latitude, and time, consisting of long
intervals of gradual build-up, punctuated by episodic breakouts of
material trapped in an equatorial disk.
\item
However, the gradual buildup and storage of angular momentum
in the circumstellar field  and gas implies that the stellar spindown rate
is likewise gradual, with no sudden jumps during intervals of breakout.
As such, ``magnetospheric emptying'' does not seem like a likely
explanation for sudden jumps in rotation rate claimed in some magnetic
hot-stars.
\item
The ``dead zone'' of closed loops surrounding the equator does
inhibit the equatorial loss of angular momentum from the near the
stellar  surface, but the net effect is merely to shunt the angular momentum flux to
a magnetic component at mid-latitudes.
The upshot is that the overall dipole scaling for total ${\dot J}$
effectively already accounts for any dead-zone reduction.

\end{itemize}

The 2-D, time-dependent models here for rotation-aligned dipole thus provide
a much richer physical picture for angular momentum loss than inferred in the
1-D, steady models for the WD67 idealization of a monopole field.
However, proper interpretation for the broad population of hot-stars
inferred to have tilted dipoles or even multipole fields will require even more
challenging 3-D MHD simulation models that explicitly allow for variations in
azimuth as well as latitude and radius.
Developing such fully 3-D MHD simulations for these cases
is thus a major focus of our planned future research.

\section*{Acknowledgments}
This work was carried out with partial support by NASA Grants
Chandra/TM7-8002X and LTSA/NNG05GC36G, and by NSF grant AST-0507581.
We thank D. Mullan and A.J. van Marle for helpful discussions, and we also thank
M. Oksala for her help in literature search.

\appendix

\section{Angular momentum flux from magnetic pole for oblique rotator}
\label{sec:appendix}

To gain insight into the relative effectiveness of angular momentum loss
for the case of non-aligned dipole,
let us extend the WD67 monopole analysis to the 1-D flow directly along
the magnetic pole of an {\em oblique dipole}, i.e., with axis now
{\em perpendicular} to the rotation.
This flow tube
has
a radial orientation,
but its areal expansion is some factor $h(r)$ faster than $r^{2}$,
so that both radial mass flux and radial field strength now vary as
$\rho v_{r} \sim B_{r} \sim 1/h r^{2}$.
Following  eqn.\ (20) of \citet{OwoudD2004},
we can approximate this expansion factor by
\beq
h(r) = \frac{r}{R_{\ast}} \,
\frac{R_{\ast} + R_{A}}{r + R_{A}}
\, ,
\label{hrdef}
\eeq
where we take their ``confinement radius'' $R_{c} = R_{A}$.

The analysis then proceeds analogously to the WD monopole case
summarized in section 2.3, leading again to an equation of the form
(\ref{jg-wd}), except that now the quantities
${\dot m} \equiv 4 \pi \rho v r^{2}$, and thus
$j_{eq} \equiv {\dot J}_{eq}/{\dot m}$, are no longer constant in radius.
Instead, in terms of the constant mass loss rate ${\dot M}$, we have
${\dot m} (r) = {\dot M}/h(r)$.
Continuity at the Alfv\'{e}n critical point again requires
$j_{eq} = \Omega R_{A}^{2}$, but this now implies an
 angular momentum loss that scales as
\beq
{\dot J}_{eq}
= {\dot M}
\frac{\Omega R_{A}^{2}}
{h_{A}}
\, ,
\eeq
where $h_{A} \equiv h(R_{A}) = (1+R_{A}/\Rstar)/2$.

Since $\rho v_{r} \sim B_{r} \sim 1/h r^{2}$ along this dipole-axis
flow, the ratio of magnetic to flow energy follows the scaling
\beq
\eta (r) \equiv \frac{B^{2}}{4 \pi \rho v_{r}^{2}}
=
\frac{ 4 \eta_{\ast} \Rstar^{2}}{h(r) r^{2}
(1-\Rstar/r)^{\beta} }
\, ,
\label{eta-dipaxis}
\eeq
where the factor 4 accounts for the fact that $\eta_{\ast}$ is defined
in terms of $B_{eq}^{2} = B_{p}^{2}/4$.
Ignoring for simplicity the radial variation of velocity
(effectively using $\beta=0$),
the Alfv\'{e}n condition $\eta(R_{A}) \equiv 1$ thus implies
$(R_{A}/\Rstar)^{2} = 4 \eta_{\ast}/h_{A}$, and so
\beq
{\dot J}_{eq}
= {\dot M} \, \Omega R_{\ast}^{2}
\frac{4 \eta_{\ast}}
{h_{A}^{2}}
= {\dot M} \, \Omega R_{\ast}^{2}
\frac{16 \eta_{\ast}}
{ \left ( 1 + R_{A}/\Rstar \right )^{2}}
\, .
\eeq
In the strong confinement limit $\eta_{\ast} \gg 1$, we find
$R_{A}/\Rstar \approx 2 \eta_{\ast}^{1/3}$, which gives
\beq
{\dot J}_{eq}
\approx {\dot M} \, \Omega R_{\ast}^{2} ~
4 \eta_{\ast}^{1/3} ~~ ; ~~ \eta_{\ast} \gg 1
\, .
\eeq
Comparing this to the monopole scaling ${\dot J} \sim \eta_{\ast}$ and
the aligned-dipole scaling ${\dot J} \sim \sqrt{\eta_{\ast}}$, we see
that this oblique dipole-axis has an angular momentum loss rate that
is weaker than either of them, scaling only as ${\dot J} \sim
\eta_{\ast}^{1/3}$.


\begin{thebibliography}{}

\bibitem[\protect\citeauthoryear{{Castor}, {Abbott} \& {Klein}}{{Castor}
  et~al.}{1975}]{Cas1975}
{Castor} J.~I.,  {Abbott} D.~C.,    {Klein} R.~I.,  1975, \apj, 195, 157

\bibitem[\protect\citeauthoryear{{Claret}}{{Claret}}{2004}]{Cla2004}
{Claret} A.,  2004, \aap, 424, 919

\bibitem[\protect\citeauthoryear{{Donati}, {Babel}, {Harries}, {Howarth},
  {Petit} \& {Semel}}{{Donati} et~al.}{2002}]{Don2002}
{Donati} J.-F.,  {Babel} J.,  {Harries} T.~J.,  {Howarth} I.~D.,  {Petit} P.,
   {Semel} M.,  2002, \mnras, 333, 55

\bibitem[\protect\citeauthoryear{{Donati}, {Howarth}, {Bouret}, {Petit},
  {Catala} \& {Landstreet}}{{Donati} et~al.}{2006}]{Don2006b}
{Donati} J.-F.,  {Howarth} I.~D.,  {Bouret} J.-C.,  {Petit} P.,  {Catala} C.,
   {Landstreet} J.,  2006, \mnras, 365, L6

\bibitem[\protect\citeauthoryear{{Friend} \& {MacGregor}}{{Friend} \&
  {MacGregor}}{1984}]{FriMac1984}
{Friend} D.~B.,  {MacGregor} K.~B.,  1984, \apj, 282, 591

\bibitem[\protect\citeauthoryear{{Groote} \& {Hunger}}{{Groote} \&
  {Hunger}}{1982}]{GroHun1982}
{Groote} D.,  {Hunger} K.,  1982, \aap, 116, 64

\bibitem[\protect\citeauthoryear{{Kawaler}}{{Kawaler}}{1988}]{Kaw1988}
{Kawaler} S.~D.,  1988, \apj, 333, 236

\bibitem[\protect\citeauthoryear{{Keppens} \& {Goedbloed}}{{Keppens} \&
  {Goedbloed}}{2000}]{KepGoe2000}
{Keppens} R.,  {Goedbloed} J.~P.,  2000, \apj, 530, 1036

\bibitem[\protect\citeauthoryear{{Kholtygin}, {Fabrika}, {Burlakova},
  {Valyavin}, {Chuntonov}, {Kudryavtsev}, {Kang}, {Yushkin} \&
  {Galazutdinov}}{{Kholtygin} et~al.}{2007}]{Kho2007}
{Kholtygin} A.~F.,  {Fabrika} S.~N.,  {Burlakova} T.~E.,  {Valyavin} G.~G.,
  {Chuntonov} G.~A.,  {Kudryavtsev} D.~O.,  {Kang} D.,  {Yushkin} M.~V.,
  {Galazutdinov} G.~A.,  2007, Astronomy Reports, 51, 920

\bibitem[\protect\citeauthoryear{{Krti{\v c}ka}, {Kub{\'a}t} \&
  {Groote}}{{Krti{\v c}ka} et~al.}{2006}]{KrtKub2006}
{Krti{\v c}ka} J.,  {Kub{\'a}t} J.,    {Groote} D.,  2006, \aap, 460, 145

\bibitem[\protect\citeauthoryear{{MacGregor} \& {Cassinelli}}{{MacGregor} \&
  {Cassinelli}}{2003}]{MacCas2003}
{MacGregor} K.~B.,  {Cassinelli} J.~P.,  2003, \apj, 586, 480

\bibitem[\protect\citeauthoryear{{MacGregor} \& {Charbonneau}}{{MacGregor} \&
  {Charbonneau}}{1999}]{MacCha1999}
{MacGregor} K.~B.,  {Charbonneau} P.,  1999, \apj, 519, 911

\bibitem[\protect\citeauthoryear{{MacGregor}, {Friend} \&
  {Gilliland}}{{MacGregor} et~al.}{1992}]{Mac1992}
{MacGregor} K.~B.,  {Friend} D.~B.,    {Gilliland} R.~L.,  1992, \aap, 256, 141

\bibitem[\protect\citeauthoryear{{Matt} \& {Balick}}{{Matt} \&
  {Balick}}{2004}]{MatBal2004}
{Matt} S.,  {Balick} B.,  2004, \apj, 615, 921

\bibitem[\protect\citeauthoryear{{Matt} \& {Pudritz}}{{Matt} \&
  {Pudritz}}{2008}]{MatPud2008}
{Matt} S.,  {Pudritz} R.~E.,  2008, \apj, 678, 1109

\bibitem[\protect\citeauthoryear{{Mestel}}{{Mestel}}{1968a}]{Mes1968a}
{Mestel} L.,  1968a, \mnras, 138, 359

\bibitem[\protect\citeauthoryear{{Mestel}}{{Mestel}}{1968b}]{Mes1968b}
{Mestel} L.,  1968b, \mnras, 140, 177

\bibitem[\protect\citeauthoryear{{Mestel}}{{Mestel}}{1984}]{Mes1984}
{Mestel} L.,  1984, in {Baliunas} S.~L.,  {Hartmann} L.,  eds, Cool Stars,
  Stellar Systems, and the Sun Vol.~193 of Lecture Notes in Physics, Berlin
  Springer Verlag, {Angular Momentum Loss During Pre-Main Sequence
  Contraction}.
pp 49--+

\bibitem[\protect\citeauthoryear{{Mestel} \& {Spruit}}{{Mestel} \&
  {Spruit}}{1987}]{MesSpr1987}
{Mestel} L.,  {Spruit} H.~C.,  1987, \mnras, 226, 57

\bibitem[\protect\citeauthoryear{{Mikul{\'a}{\v s}ek} \& {et
  al.}}{{Mikul{\'a}{\v s}ek} \& {et al.}}{2008}]{Mik2008}
{Mikul{\'a}{\v s}ek} Z.,  {et al.} 2008, \aap, 485, 585

\bibitem[\protect\citeauthoryear{{Mullan} \& {MacDonald}}{{Mullan} \&
  {MacDonald}}{2001}]{MulMac2001}
{Mullan} D.~J.,  {MacDonald} J.,  2001, \apj, 559, 353

\bibitem[\protect\citeauthoryear{{Mullan} \& {MacDonald}}{{Mullan} \&
  {MacDonald}}{2005}]{MulMac2005}
{Mullan} D.~J.,  {MacDonald} J.,  2005, \mnras, 356, 1139

\bibitem[\protect\citeauthoryear{{Neiner}, {Geers}, {Henrichs}, {Floquet},
  {Fr{\'e}mat}, {Hubert}, {Preuss} \& {Wiersema}}{{Neiner}
  et~al.}{2003}]{Nei2003}
{Neiner} C.,  {Geers} V.~C.,  {Henrichs} H.~F.,  {Floquet} M.,  {Fr{\'e}mat}
  Y.,  {Hubert} A.-M.,  {Preuss} O.,    {Wiersema} K.,  2003, \aap, 406, 1019

\bibitem[\protect\citeauthoryear{{Okamoto}}{{Okamoto}}{1974}]{Oka1974}
{Okamoto} I.,  1974, \mnras, 166, 683

\bibitem[\protect\citeauthoryear{{Owocki} \& {ud-Doula}}{{Owocki} \&
  {ud-Doula}}{2004}]{OwoudD2004}
{Owocki} S.~P.,  {ud-Doula} A.,  2004, \apj, 600, 1004

\bibitem[\protect\citeauthoryear{{Sakurai}}{{Sakurai}}{1985}]{Sak1985}
{Sakurai} T.,  1985, \aap, 152, 121

\bibitem[\protect\citeauthoryear{{Smith} \& {Bohlender}}{{Smith} \&
  {Bohlender}}{2007}]{SmiBoh2007}
{Smith} M.~A.,  {Bohlender} D.~A.,  2007, \aap, 466, 675

\bibitem[\protect\citeauthoryear{{St{\c e}pie{\'n}}}{{St{\c
  e}pie{\'n}}}{1998}]{Ste1998}
{St{\c e}pie{\'n}} K.,  1998, \aap, 337, 754

\bibitem[\protect\citeauthoryear{{Stone} \& {Norman}}{{Stone} \&
  {Norman}}{1992}]{StoNor1992}
{Stone} J.~M.,  {Norman} M.~L.,  1992, \apjs, 80, 753

\bibitem[\protect\citeauthoryear{{Suess} \& {Nerney}}{{Suess} \&
  {Nerney}}{1975}]{SueNer1975}
{Suess} S.~T.,  {Nerney} S.~F.,  1975, \solphys, 40, 487

\bibitem[\protect\citeauthoryear{{Thompson} \& {Landstreet}}{{Thompson} \&
  {Landstreet}}{1985}]{ThoLan1985}
{Thompson} I.~B.,  {Landstreet} J.~D.,  1985, \apjl, 289, L9

\bibitem[\protect\citeauthoryear{{Tout} \& {Pringle}}{{Tout} \&
  {Pringle}}{1992}]{TouPri1992}
{Tout} C.~A.,  {Pringle} J.~E.,  1992, \mnras, 256, 269

\bibitem[\protect\citeauthoryear{{Trigilio}, {Leto}, {Umana}, {Buemi} \&
  {Leone}}{{Trigilio} et~al.}{2008}]{Tri2008}
{Trigilio} C.,  {Leto} P.,  {Umana} G.,  {Buemi} C.~S.,    {Leone} F.,  2008,
  \mnras, 384, 1437

\bibitem[\protect\citeauthoryear{{ud-Doula} \& {Owocki}}{{ud-Doula} \&
  {Owocki}}{2002}]{udDOwo2002}
{ud-Doula} A.,  {Owocki} S.~P.,  2002, \apj, 576, 413

\bibitem[\protect\citeauthoryear{{ud-Doula}, {Owocki} \& {Townsend}}{{ud-Doula}
  et~al.}{2008}]{udDOwo2008}
{ud-Doula} A.,  {Owocki} S.~P.,    {Townsend} R.~H.~D.,  2008, \mnras, 385, 97

\bibitem[\protect\citeauthoryear{{Washimi} \& {Shibata}}{{Washimi} \&
  {Shibata}}{1993}]{WasShi1993}
{Washimi} H.,  {Shibata} S.,  1993, \mnras, 262, 936

\bibitem[\protect\citeauthoryear{{Weber} \& {Davis}}{{Weber} \&
  {Davis}}{1967}]{WebDav1967}
{Weber} E.~J.,  {Davis} L.~J.,  1967, \apj, 148, 217

\end{thebibliography}
\end{document}